\documentstyle[preprint,aps]{revtex}
\newcommand\eqn[1]{\begin{equation}#1\end{equation}}
\newcommand\eqa[1]{\begin{eqnarray}#1\end{eqnarray}}

\newcommand\half{\frac{1}{2}}

\title{
Analysis of ${\bf \Delta}${\bf I = 2} Staggering\\
in Nuclear Rotational Spectra
}

\author{
K. Hara\\
Physik-Department, Technische Universit\"{a}t M\"{u}nchen\\
D-85747 Garching bei M\"{u}nchen, Germany\\
and\\
G. A. Lalazissis\\
Department of Theoretical Physics, Aristotle University of
Thessaloniki\\ GR-54006 Thessaloniki, Greece
}

\begin{document}

\maketitle

\begin{abstract}

A method is proposed and tested for the analysis of $\Delta I=2$
staggering observed in nuclear rotational bands. We examine six super-
and hyper-deformed bands, among which that of $^{149}$Gd and possibly of
$^{147}$Gd seem to exhibit real staggering. However, we emphasize that
the presence of staggering may not necessarily imply the occurrence of
bifurcation. It is also shown that a similar staggering seen in normally
deformed bands is a manifestation of band crossings. A more extensive
analysis is planned.

\end{abstract}

\newpage


Recently, a striking feature of $\Delta I=2$ staggering in rotational
spectra has been reported for several super- and hyper-deformed bands
[1-7] attracting much attention and interest in the nuclear physics
community. As a result, it has become a most frequently debated subject
and a considerable amount of effort has been spent on understanding its
physical implication based on various theoretical ideas [8-14].
Nevertheless, definite conclusions have not yet been reached until
present time. On the other hand, a similar staggering can also be
observed in molecular rotational spectra, whose underlying mechanism is
not known either. In particular, for diatomic molecules [15-22], the
occurrence of staggering is definitely not due to the presence of $C_4$
type symmetry of the system because of their dumb-bell structure. In
both nuclear and molecular cases, one presents the experimental data in
terms of a 5-point formula (which is denoted as $\Delta E_\gamma$ in
Ref.[2])
\eqn{
\Delta^{(5)}E_2(I) \equiv {1 \over {16}} \big[6E_2(I)
-4E_2(I+2)-4E_2(I-2)+E_2(I+4)+E_2(I-4) \big]
} 
where we have introduced the notation for an n-point formula $\Delta^{(n)}
E_{\Delta I}(I)$ with respect to
\eqn{
E_{\Delta I}(I) \equiv E(I) - E(I-\Delta I).
} 
It is the regularity of the result obtained by this formula that one is
interested in. It may be instructive to note that, although there is a
close similarity between the observation in nuclear and molecular
systems, their dynamical laws and the energy scales are totally
different except for the kinematical aspect that both of them are
``rotating''. This seems to suggest that the phenomenon of staggering is
rather independent of characteristic properties of actual systems. The
purpose of the present work is to gain more information and insight
about the nature of staggering by analyzing experimental data.

A typical example of the $\Delta I=2$ staggering in nuclear rotational
spectra is presented in Fig. 1 for a super-deformed band of 
the nucleus $^{149}$Gd \cite {gd149a}.

\medskip\bigskip\hrule\bigskip
\hfil Fig. 1 \hfil
\bigskip\hrule\bigskip\medskip

\noindent
It is usual to interpret such a $\Delta I=2$ staggering as the
occurrence of a $\Delta I=4$ bifurcation in the corresponding spectrum.
However, we stress that it may not necessarily be true. In fact, one
notices a remarkable fact that the same spectrum shows also a $\Delta
I=4$ staggering if one applies the 5-point formula with $\Delta I=4$
\cite {george}
\eqn{
\Delta^{(5)}E_4(I) \equiv {1 \over {16}} \big[6E_4(I)
-4E_4(I+4)-4E_4(I-4)+E_4(I+8)+E_4(I-8) \big].
} 
Note that $E_4(I)$ is obtained from the $\Delta I=2$ transition energy
$E_2(I)$ using the relation
\[
E_4(I)=E(I)-E(I-4)=E_2(I)+E_2(I-2).
\]
Figs. 2 shows the 5-point formula (3) resulting from the same data
as Figs. 1.

\medskip\bigskip\hrule\bigskip
\hfil Fig. 2 \hfil
\bigskip\hrule\bigskip\medskip

\noindent
According to the usual interpretation, one would conclude from Figs. 2
that there exists a $\Delta I=8$ bifurcation. Similarly, for a spectrum
measured over a sufficiently wide range of spins as in the case of many
molecular bands, one can observe a $\Delta I=6$ staggering in the
quantity $\Delta^{(5)}E_6(I)$, from which one might conclude the
existence of a $\Delta I=12$ bifurcation \cite {george}. We would
consider it rather questionable to regard such a result as a finger
print of bifurcations. In the same token, the $\Delta I=2$ staggering
might not necessarily imply the presence of a $\Delta I=4$ bifurcation.
Although the bifurcation may be a possible interpretation, it has to be
proved (or disproved) by further theoretical investigations.
Accordingly, it seems necessary to gain more information and insight
about the nature of the staggering from the existing data and this is
what we aim at in the present work.

A large number of similar staggering can be found also in the yrast
spectra of normally deformed doubly even rare-earth nuclei. Let us take
the nucleus $^{160}$Yb \cite {yb} as an example. Figs. 3 and 4 show the
results of 5-point formulas (1) and (3), respectively.

\medskip\bigskip\hrule\bigskip
\hfil Fig. 3 \hfil Fig. 4 \hfil
\bigskip\hrule\bigskip\medskip

\noindent
Although one sees both $\Delta I=2$ and $\Delta I=4$ staggering in these
figures, no one would believe that the yrast spectrum of $^{160}$Yb
possesses any kind of bifurcation. On the one hand, the formulas (1)
and (3) are proportional roughly to the ``fifth'' order derivative of
the energy $E(I)$ in finite differences. On the other hand, it is known
that there are crossings between the g- and s-band around $I=12$ and
between the s- and a 4-quasiparticle band around $I=26$ in this nucleus
\cite {hs1}. This fact gives us a hint about a possible origin of the
staggering at least for the present case. Indeed, if there occurs a band
crossing, the spectrum will have a kink at the crossing point so that
its derivatives become discontinuous. Because of the use of finite
differences, however, this discontinuity propagates over other spin
values and appears as a staggering over a certain spin range around the
crossing point. Obviously, this is a localized phenomenon centered at
the crossing point but, if more than one band crossing takes place
sequentially, the staggering can extend over a wider spin region. This
is what we actually see in Figs. 3 and 4.

In order to make the mechanism qualitatively described above more
quantitative and clearly visible, let us tailor a simple schematic model
in which two band crossings take place among three unperturbed bands,
which is essentially an extension of a two-band crossing model discussed
in \cite{hs}. Their rotational energies will be assumed as
\eqa{
E_0&=&A_0I(I+1) \nonumber \\
E_1&=&A_1\big[I(I+1)-2B_1(I-I_1)\big]+C_1 \\
E_2&=&A_2\big[I(I+1)-2B_2(I-I_2)\big]+C_2 \nonumber
} 
with $A_0>A_1>A_2$. The band 0 and 1 are assumed to cross each other at
$I=I_1$ and band 1 and 2 at $I=I_2$ ($I_1<I_2$). The quantity $B_1$
($B_2$) controls the crossing angle between band 0 and 1 (1 and 2) while
$C_1$ and $C_2$ are determined by the crossing conditions such that
$E_0=E_1$ at $I=I_1$ and $E_1=E_2$ at $I=I_2$, so that they are given by
\eqn{
C_1=(A_0-A_1)I_1(I_1+1),\ \ C_2=(A_1-A_2)I_2(I_2+1)-2A_1B_1(I_2-I_1)
+C_1.
} 
For simplicity, we assume that only the lowest two bands couple with
each other as the band higher than the lowest two is unimportant for the
yrast energy in the crossing region. The coupling strength will be taken
to be spin independent. The yrast energy of the system thus becomes
\eqn{
E(I)=E-\sqrt{D^2+V^2}
} 
where the quantities $E$, $D$ and $V$ are defined by
\eqn{
E=\half(E_0+E_1),\ \ D=\half(E_0-E_1)\ \ \mbox{and}\ \ V=V_{01}
\ \ \mbox{if}\ \ E_0 \le E_2
} 
and
\eqn{
E=\half(E_1+E_2),\ \ D=\half(E_1-E_2)\ \ \mbox{and}\ \ V=V_{12}
\ \ \mbox{if}\ \ E_0 > E_2.
} 
To visualize the situation clearly, we present a diagram for the
rotational energies of these bands (lines) together with the
yrast energy (dots) in Fig. 5.

\medskip\bigskip\hrule\bigskip
\hfil Fig. 5 \hfil
\bigskip\hrule\bigskip\medskip

\noindent
The yrast spectrum has a kink at a crossing point. We note that the
sharpness of a kink determines the amplitude of the resulting
staggering. Obviously, the larger the crossing angle and/or the weaker
the coupling strength, the sharper the kink making the amplitude of the
staggering larger. In fact, if the band coupling is switched off, the
yrast energy (6) becomes $E(I)=E-|D|$ which leads to the sharpest
possible kink for a given crossing angle. In this limiting case, the
discontinuity of derivatives of $E(I)$ at a crossing point arises from
the term $|D|$ because $D$ changes sign at each crossing point.

\medskip\bigskip\hrule\bigskip
\hfil Fig. 6 \hfil Fig. 7 \hfil
\bigskip\hrule\bigskip\medskip

\noindent
Figs. 6 and 7 show respectively the case of vanishing and non-vanishing
coupling strength for the yrast energy (6) using the formula (1), which
produces a $\Delta I=2$ staggering. Needless to say, the formula (3)
will produce a $\Delta I=4$ staggering if one uses it in place of the
formula (1). The feature of a staggering depends sensitively on the
crossing angle, the relative position of the crossing points and the
strength of the coupling. One can simulate and study possible features
of staggering by changing the parameters of the model. It should be
clearly stated that the staggering in question is caused by the use of a
5-point formula and is not due to a physical effect. It is merely a
manifestation of band crossings which produce kinks in the spectrum. For
the analysis of data, one should avoid using formulas such as Eqs. (1)
and (3) which are not free from the effect of band crossings. Such an
effect has to be first removed if one wants to see the presence of real
(physical) staggering. On the one hand, what we are ultimately
interested in is the actual behavior of the $\Delta I=2$ transition
energy $E_2(I)$. On the other hand, this quantity is a globally
increasing function of spin extending from some 100 keV to a value well
beyond 1 MeV, so that its fine variations of less than 1 keV are
invisible in the plot of $E_2(I)$. However, this is only a matter of
proper scaling. We can circumvent it by subtracting a smoothly
increasing part from the measured transition energy $E_2(I)$.

For this purpose, let us define a 1-point formula by the expression
\eqn{
\Delta^{(1)}E_2(I) \equiv E_2(I)-a-bI-cI^2-dI^3.
} 
The coefficients $a$, $b$, $c$ and $d$ are determined by minimizing the
function
\eqn{
f(a,b,c,d)=\sum_I^{\Delta I=2} \big[\Delta^{(1)}E_2(I)\big]^2
} 
with respect to $a$, $b$, $c$ and $d$. The summation over spin $I$ in
Eq. (10) is taken in step of $\Delta I=2$. If the absolute spins are not
known as in the case of a super- or hyper-deformed band, we may measure
them with respect to a (unknown) reference spin $I_{ref}$, which means
that we simply take $I=2,4,\cdots$. The minimization conditions
\[
{\partial f(a,b,c,d)\over{\partial a}}=
{\partial f(a,b,c,d)\over{\partial b}}=
{\partial f(a,b,c,d)\over{\partial c}}=
{\partial f(a,b,c,d)\over {\partial d}}=0
\]
lead to a set of linear equations for $a$, $b$, $c$ and $d$
\eqa{
S_0a+S_1b+S_2c+S_3d&=&F_0 \nonumber \\
S_1a+S_2b+S_3c+S_4d&=&F_1 \\
S_2a+S_3b+S_4c+S_5d&=&F_2 \nonumber \\
S_3a+S_4b+S_5c+S_6d&=&F_3 \nonumber
} 
where the coefficients in these equations\footnote{The determinant of
Eq. (11) is ill-conditioned. First eliminate $d$ by using the fourth
equation and then solve the resulting linear equations for $a$, $b$ and
$c$ for the sake of numerical stability.} are defined by
\eqn{
S_n=\sum_I^{\Delta I=2} I^n,\ \ F_n=\sum_I^{\Delta I=2} I^n E_2(I).
} 
The subtracting part $a+bI+cI^2+dI^3$ is thus nothing other than the
$\chi$-square fit of $E_2(I)$. For the schematic model presented above
as well as for a normally deformed band, subtracting a linear term
$a+bI$ is sufficient. However, for a super-/hyper-deformed and molecular
band, the spectrum will behave globally as
\eqn{
E(I) \approx AI(I+1)+B\big[I(I+1)\big]^2,
} 
so that inclusion of higher order terms of $I$ may be significant. We
note that, depending on the sign of the quantity $B$ in Eq. (13), the
effective moment of inertia will decrease ($B>0$, a stretched rotational
spectrum) or increase ($B<0$, a compressed rotational spectrum) as a
function of spin. Both cases are possible for super-/hyper-deformed
bands while molecular bands belong to the latter. We will come back to
this point later.

The 1-point formula (9) would not change the staggering feature of
$E_2(I)$, if any. The quantity $\Delta^{(1)}E_2(I)$ represents a
deviation from the mean (smooth) behavior of $E_2(I)$ and changes its
sign by construction which should not be confused with regular
oscillations. In particular, a sudden decrease of the value of
$\Delta^{(1)}E_2(I)$ implies that there is a band crossing. If the
regular oscillations which are present in $\Delta^{(5)}E_2(I)$ disappear
in $\Delta^{(1)}E_2(I)$, it means that this staggering is produced by
band crossings or more generally by kinks in $\Delta^{(1)} E_2(I)$. It
should be stressed that even a weak kink in $\Delta^{(1)} E_2(I)$ will
cause a discontinuity of its derivatives and this will produce a
staggering in $\Delta^{(5)} E_2(I)$. In other words, the 5-point formula
(1) is so fragile that it may well happen that the same band measured
at different laboratories might exhibit different staggering features
due to different experimental uncertainties. In contrast, the 1-point
formula (9) is robust and reliable. On the other hand, if the quantity
$\Delta^{(1)} E_2(I)$ exhibits regular oscillations, there can be two
possible reasons, the occurrence of either a real (physical) staggering
or successive band crossings that take place closely one after the
other. One cannot distinguish these two cases from each other without
going into a detailed theoretical analysis of the system. We now present
results of the 1-point formula (9).

\medskip\bigskip\hrule\bigskip
\hfil Fig. 8 \hfil Fig. 9 \hfil
\bigskip\hrule\bigskip\medskip

Figs. 8 and 9 show the results of the 1-point formula (9) applied to
the schematic model (cf. Fig. 7) and the nucleus $^{160}$Yb (cf. Fig.
4), respectively. These figures exhibit no regular oscillation and thus
indicate correctly that the staggering seen in Figs. 7 and 4 stem from
band crossings. A weak (third) band crossing is seen in the nucleus
$^{160}$Yb in the highest spin region. In general, we can state that the
mechanism leading to the staggering in normally deformed bands can be
attributed to such band crossings. These examples are however just to
show how the 1-point formula works. In what follows, we turn to the
analysis of super-/hyper-deformed bands.

\medskip\bigskip\hrule\bigskip
\hfil Fig. 10 \hfil
\bigskip\hrule\bigskip\medskip

Fig. 10 shows the 1-point formula (9) applied to the super-deformed band
of the nucleus $^{149}$Gd (cf. Fig. 1). In this figure, one sees regular
oscillations in the spin region 16 -- 36 (relative to a unknown
reference spin $I_{ref}$). It is most likely that they do not stem from
band crossings. The reason will be given below in connection with the
(dynamic) moment of inertia.

The dynamic moment of inertia (a 2-point formula)
\eqn{
\Theta^{(2)} = {4 \over {\Delta^{(2)}E_2(I)}},
\ \ \Delta^{(2)}E_2(I) \equiv E_2(I+2)-E_2(I)
} 
can provide us with some information about the staggering depending on
the situation. As mentioned before, the quantity $B$ in Eq. (13) can be
positive (a stretched rotational spectrum) or negative (a compressed
rotational spectrum) in super- and hyper-deformed bands corresponding to
globally decreasing (e.g. $^{149}$Gd and $^{154}$Er) or globally
increasing (e.g. $^{191}$Hg and $^{194}$Hg) moment of inertia, although
there is the third case in which the moment of inertia oscillates
about a certain value (e.g. $^{147}$Gd and $^{195}$Pb). If the moment of
inertia decreases, one may exclude the occurrence of band crossing. In
fact, it should increase if there occurs a band crossing because the
moment of inertia of the crossing band has to be larger than that of the
crossed band. Figs. 11 and 12 compare the dynamic moments of inertia for
$^{149}$Gd \cite {gd149a} and $^{194}$Hg \cite {hg194}, respectively.

\medskip\bigskip\hrule\bigskip
\hfil Fig. 11 \hfil Fig. 12 \hfil
\bigskip\hrule\bigskip\medskip

\noindent
One can exclude band crossing in the case of the nucleus $^{149}$Gd as
its moment of inertia decreases (globally) with spin. One may thus
conclude that the staggering in this nucleus is most likely a real
(physical) one. However, for the nucleus $^{194}$Hg, one cannot exclude
possible band crossing as its moment of inertia increases. Under such a
circumstance, it is most interesting to compare the 5-point formula (1)
with the 1-point formula (9). This is done in Figs. 13 and 14.

\medskip\bigskip\hrule\bigskip
\hfil Fig. 13 \hfil Fig. 14 \hfil
\bigskip\hrule\bigskip\medskip

\noindent
Although one may regard the 5-point formula for $^{194}$Hg practically
as zero because of large error bars, it shows a staggering if one takes
the most probable values of data. On the other hand, the 1-point formula
for this nucleus is outside the error bars and clearly shows irregular
kinks\footnote{We note in passing that a staggering can be regarded as a
series  of ``regular kinks'' due to large and small function values that
occur alternatingly.}. It is clear that the staggering of $\Delta^{(5)}
E_2(I)$ for $^{194}$Hg stems from irregular kinks which lead to
discontinuities in the derivative of $\Delta^{(1)}E_2(I)$. Therefore,
the staggering in question is not a real one. The result for the nucleus
$^{191}$Hg \cite {hg191} is similar. On the other hand, the moment of
inertia of $^{147}$Gd \cite {gd147} and $^{195}$Pb \cite {pb195}
oscillate about some values. The 1-point formula for $^{147}$Gd shows a
staggering which is not so regular as in the case of $^{149}$Gd while
that of $^{195}$Pb shows mostly irregular kinks although a very short
oscillation can be seen if one discards error bars. Finally, the moment
of inertia of the nucleus $^{154}$Er \cite {er154} decreases as in the
case of $^{149}$Gd. However, the 1-point formula shows only irregular
kinks so that the staggering in this nucleus is not a real one either.
Fig. 15 summarizes the results of the 1-point formula quoted in these
discussions.

\medskip\bigskip\hrule\bigskip
\hfil Fig. 15 \hfil
\bigskip\hrule\bigskip\medskip

In Fig. 15, all error bars are removed for the sake of clarity but we
remark that one may not conclude the presence of oscillations if the
amplitude of the staggering is small because of large error bars. This
applies in particular to the nucleus $^{191}$Hg and $^{195}$Pb. Among
six nuclei studied in the present work, the nucleus $^{149}$Gd and
possibly $^{147}$Gd remain as candidates of exhibiting real staggering,
although a possibility of band crossings cannot be excluded in the
latter. However, as we stressed before, whether a staggering implies
bifurcation of the corresponding spectrum or not has to be investigated
in future studies. On the other hand, kinks that occur in the 1-point
formula inherently produce oscillations in the 5-point formula, so that
a staggering of such an origin should not be accepted as a real
(physical) one, although there may still be a room for debating whether
or not random kinks in measured transition energies have physical
significance to be studied in more detail. In any case, irregular kinks
should be understood differently from a regular staggering. We mention
in this connection that, theoretically, an irregularity would be more
difficult to understand than a regularity.

To summarize, we have attempted to obtain as much information as
possible from (some) existing data to gain more insight to the nature of
observed staggering. In the first place, we called attention to the fact
that the use of 5-point formulas (1) and (3) always produce a $\Delta
I=2$ and $\Delta I=4$ staggering, respectively. This fact warns of the
danger of concluding the existence of bifurcations simply from the
presence of staggering patterns. We emphasize that, even if a staggering
turns out to be a real one, it may or may not imply a bifurcation. At
present, its physical interpretation is not well established and thus
has to be still sought in future theoretical studies. We also called
attention to the fact that band crossings can lead to a staggering if
one uses a multi-point formula. For normally deformed bands, this is the
origin of the staggering.

Most generally speaking, the feature of a staggering depends on the
multi-point formula one uses for the presentation of data as it
originates from (regular as well as irregular) kinks that occur in
measured transition energies. In fact, a 3-point formula
$$
\Delta^{(3)}E_2(I)\equiv E_2(I)-{1 \over 2}\big[E_2(I+2)+E_2(I-2)\big]
$$
will produce less number of oscillations than the 5-point formula \cite
{hs}. To avoid such a formula-dependence as well as the effect of band
crossings, we proposed to use the 1-point formula (9) which measures a
deviation from the smooth behavior. It provides us with more direct
information of the transition energy than multi-point formulas.

The following table summarizes the result of six super- and
hyper-deformed bands studied in the present work. The question mark in
the last column for the nucleus $^{147}$Gd is to indicate that the item
is not certain but probable.

\bigskip\bigskip
\centerline{Table: Classification of super- and hyper-deformed bands}
\medskip
\begin{center}
\begin{tabular}{|c||c|c|c|} \hline
~Nucleus~ & ~5-Point Formula~ & ~Dyn. Mom. Inertia~ & ~1-Point Formula~
\\ \hline \hline
$^{147}$Gd & staggering & oscillating & ~staggering? \\ \hline
$^{149}$Gd & staggering & decreasing  &  staggering  \\ \hline
$^{154}$Er & staggering & decreasing  & irreg. kinks \\ \hline
$^{191}$Hg & staggering & increasing  & irreg. kinks \\ \hline
$^{194}$Hg & staggering & increasing  & irreg. kinks \\ \hline
$^{195}$Pb & staggering & oscillating & irreg. kinks \\ \hline
\end{tabular}
\end{center}
\bigskip\bigskip

We stress that, before attempting a theoretical analysis of an observed
staggering, one should make sure whether the staggering in question is a
real one or simply due to irregular kinks. Otherwise, there is a danger
that one may be dealing with an object which does not exist in reality
and this may lead theoretical investigations in a wrong direction. The
proposed 1-point formula is proved to be a useful tool for filtering out
such cases, although it is quite possible that the present method of
analysis may still have to be improved. For example, one could further
subtract a fourth order term in the 1-point formula. In fact, Fig. 15
suggests the presence of such a global behavior. We shall study this in
the future when we carry out a more extensive analysis of the existing
data on super- and hyper-deformed bands.

We have done a similar analysis for molecular rotational bands and
confirmed also that staggering does not necessarily imply bifurcation.
The conclusion has been quite similar to the nuclear case presented
here. The result will be reported elsewhere.

Finally, we should like to mention that, with minor modifications
(taking $\Delta I=1$ instead of 2 and subtracting only a first order
polynomial $a+bI$ instead of the third order one), the proposed 1-point
formula is applicable to the usual signature dependent spectrum of a
normally deformed band exhibiting a genuine $\Delta I=1$ staggering
(signature dependence) which leads to the signature splitting (a $\Delta
I=2$ bifurcation). In fact, it might be quite practical for the analysis
of odd proton nuclei and in particular of doubly odd nuclei in which the
signature dependence is in most cases extremely delicate and is not
distinctly discernible if one plots the $\Delta I=1$ transition energy
$E(I)-E(I-1)$ itself. We remark that, in these nuclei, a bifurcation
(signature splitting) and a band crossing (which may cause a signature
inversion) often occur simultaneously [26, 27].

In the meantime, we have noticed that a (two-) band crossing model
similar to our (three-) band crossing model Eq. (4) is used in a
recently published work [29], in which the authors investigate the
$\Delta I=2$ staggering (mainly of the normally deformed bands) in terms
of various multi-point formulas.

\bigskip\bigskip
One of the authors (G. A. L.) acknowledges a support from the European
Union under the contract TMB-EU/ERB FMBCICT-950216.


\newpage
\parindent=2truecm
\centerline{\large FIGURE CAPTIONS}
\begin{description}

\item[Fig. ~1:] $\Delta I=2$ staggering obtained by the 5-point formula
Eq. (1) for a super-deformed band of the nucleus $^{149}$Gd

\item[Fig. ~2:] $\Delta I=4$ staggering obtained by the 5-point formula
Eq. (3) for a super-deformed band of the nucleus $^{149}$Gd

\item[Fig. ~3:] $\Delta I=2$ staggering obtained by the 5-point formula
Eq. (1) for the normally deformed band of the nucleus $^{160}$Yb

\item[Fig. ~4:] $\Delta I=4$ staggering obtained by the 5-point formula
Eq. (3) for the normally deformed band of the nucleus $^{160}$Yb

\item[Fig. ~5:] Band energies in a schematic band crossing model with
three bands 0, 1 and 2 in which the bands 0 and 1 (1 and 2) cross at
$I=12~(26)$

\item[Fig. ~6:] $\Delta I=2$ staggering obtained by the 5-point formula
Eq. (1) for the schematic band crossing model (band coupling switched
off)

\item[Fig. ~7:] $\Delta I=2$ staggering obtained by the 5-point formula
Eq. (1) for the schematic band crossing model (band coupling switched
on)

\item[Fig. ~8:] The 1-point formula Eq. (9) for the schematic model, cf.
Fig. 7

\item[Fig. ~9:] The 1-point formula Eq. (9) for the nucleus $^{160}$Yb,
cf. Fig. 4

\item[Fig. 10:] The 1-point formula Eq. (9) for the nucleus $^{149}$Gd,
cf. Fig. 1

\item[Fig. 11:] The dynamic moment of inertia Eq. (14) for the nucleus
$^{149}$Gd

\item[Fig. 12:] The dynamic moment of inertia Eq. (14) for the nucleus
$^{194}$Hg

\item[Fig. 13:] The 5-point formula Eq. (1) for the nucleus $^{194}$Hg,
cf. Fig. 14

\item[Fig. 14:] The 1-point formula Eq. (9) for the nucleus $^{194}$Hg,
cf. Fig. 13

\item[Fig. 15:] Results of the 1-point formula Eq. (9) for six
super-/hyper-deformed bands

\end{description}

\end{document}